**UNIVERSITAT ROVIRA I VIRGILI**
Escola Tècnica Superior d'Enginyeria Química
Departament d'Enginyeria Quimica

**Johann Faccelo**
**Osma Cruz**

# *Banana skin: a novel material for a low-cost production of laccase*

GRADUATE STUDIES IN CHEMICAL, ENVIRONMENTAL
AND PROCESS ENGINEERING
MASTER THESIS, 2007

This work was done under the supervision of:

Susana Rodríguez Couto, PhD;
Ramon y Cajal Research Professor, Department of Chemical Engineering - Universitat Rovira i Virgili

José Luis Toca Herera, PhD;
Group Leader - Ramon y Cajal Professor, Biosurfaces Unit CICbiomaGUNE

Tarragona, 3rd of July of 2007

The jurors:

**President**
Ioanis Katakis, PhD;
Associate Professor
Department of Chemical Engineering - Universitat Rovira i Virgili

Francesc Castells, PhD;
Professor
Department of Chemical Engineering - Universitat Rovira i Virgili

Carme Güell, PhD;
Associate Professor
Department of Chemical Engineering - Universitat Rovira i Virgili

## Acknowledgments

First, I want to thank my supervisors Susana Rodríguez-Couto and Jose Luis Toca-Herrera. The URV for the financial support during the master and research, and my group for comprehension and patience. I also want to thank my friends and family, who are the key of my life and work here.

# Index

## Index of figures

## Index of tables

* Figures 8-12 and Table 1 were published in Osma et al. 2006 [41].

# 1 Introduction

Laccases (benzenodiol: oxygen oxidoreductases; EC 1.10.3.2) are multicopper oxidases of wide substrate specificity mainly found in white-rot fungi, which are the only microorganisms able to degrade the whole wood components, but they are also expressed in bacteria and higher plants.

Laccases carry out one-electron oxidation and reduce molecular oxygen to water. The biotechnological importance of this enzyme lies in its ability to oxidize both phenolic and non-phenolic lignin-related compounds [1, 2] as well as highly recalcitrant environmental pollutants [3, 4]. This type of degradation is expected to be very useful for the treatment of wastewater and dye degradation [5-7], especially for the textile and the paper and pulp industry. The catalytic cycle of the laccase (with four cupper atoms), described by Yaropolov et al. in 1994 [8], showed the reduction of molecular oxygen into two water molecules and the catalytic oxidation of aromatic electron-donor compounds (AH) into aromatic radicals (A·) (Figure 1).

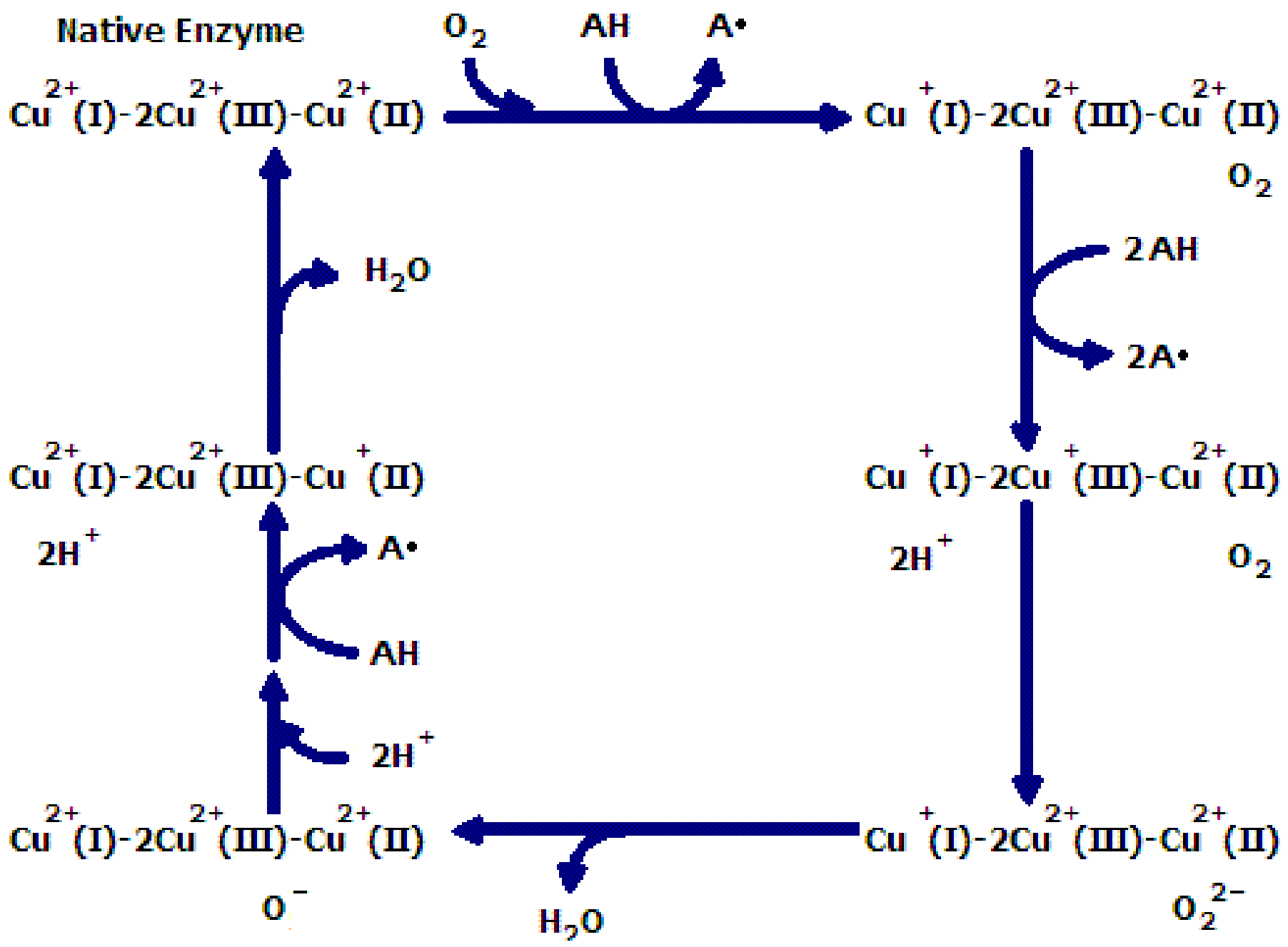

**Figure 1.** Catalytic cycle of four-copper laccase. (extracted from Yaropolov et al. 1994 [8])

The reported redox potentials of laccases (0.4 - 0.8V respect to a normal hydrogen electrode) are lower than those of non-phenolic compounds, so these enzymes cannot oxidize such substances. However, it has been shown that in the presence of small molecules capable to act as electron transfer mediators, called redox mediators, laccases are also able to oxidize non-phenolic structures [9,10] expanding the range of compounds that can be oxidized by these enzymes. The mechanism by which redox mediators play a role in laccase-catalyzed reactions is now well characterized. When a substrate is oxidized by a laccase the redox mediator (M) forms cation radicals (M+) (short-lived intermediates), which co-oxidize the substrate (S) (Figure 2). These cation radicals can be formed by two mechanisms: the redox mediator can perform either a one-electron oxidation of the substrate to a radical cation [11-13]; or it abstracts an H-atom from the substrate converting it into a radical [14,15].

Redox-mediated laccase catalysis has been used in a wide range of applications like pulp delignification, polycyclic aromatic hydrogen degradation, pesticide or insecticide degradation and organic synthesis [16]. Claus et al. (2002) [17] found that the system laccase plus mediator enhanced dye decoloration and some dyes resistant to laccase degradation were decolorized. This is a very important feature for the degradation of highly recalcitrant environmental pollutants, and increases the interest on laccases for the treatment of wastewater and different types of aromatic or lignin-related compounds.

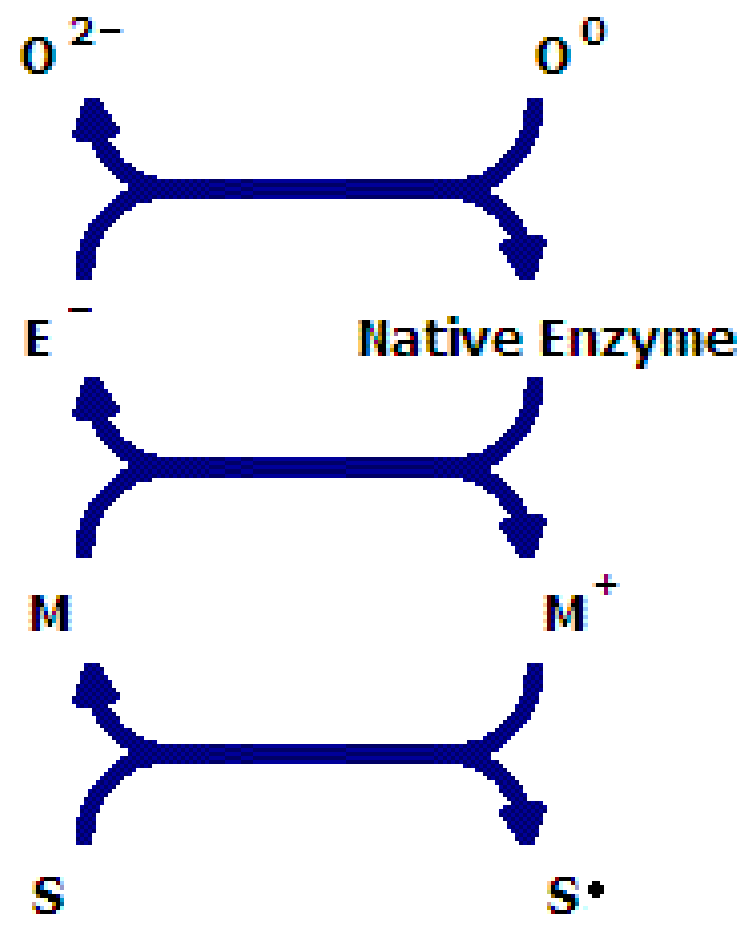

**Figure 2.** Catalytic process of a substrate (S) by laccase (E) and a redox mediator (M) (Adapted from Bourbonnais et al. 1998 [11])

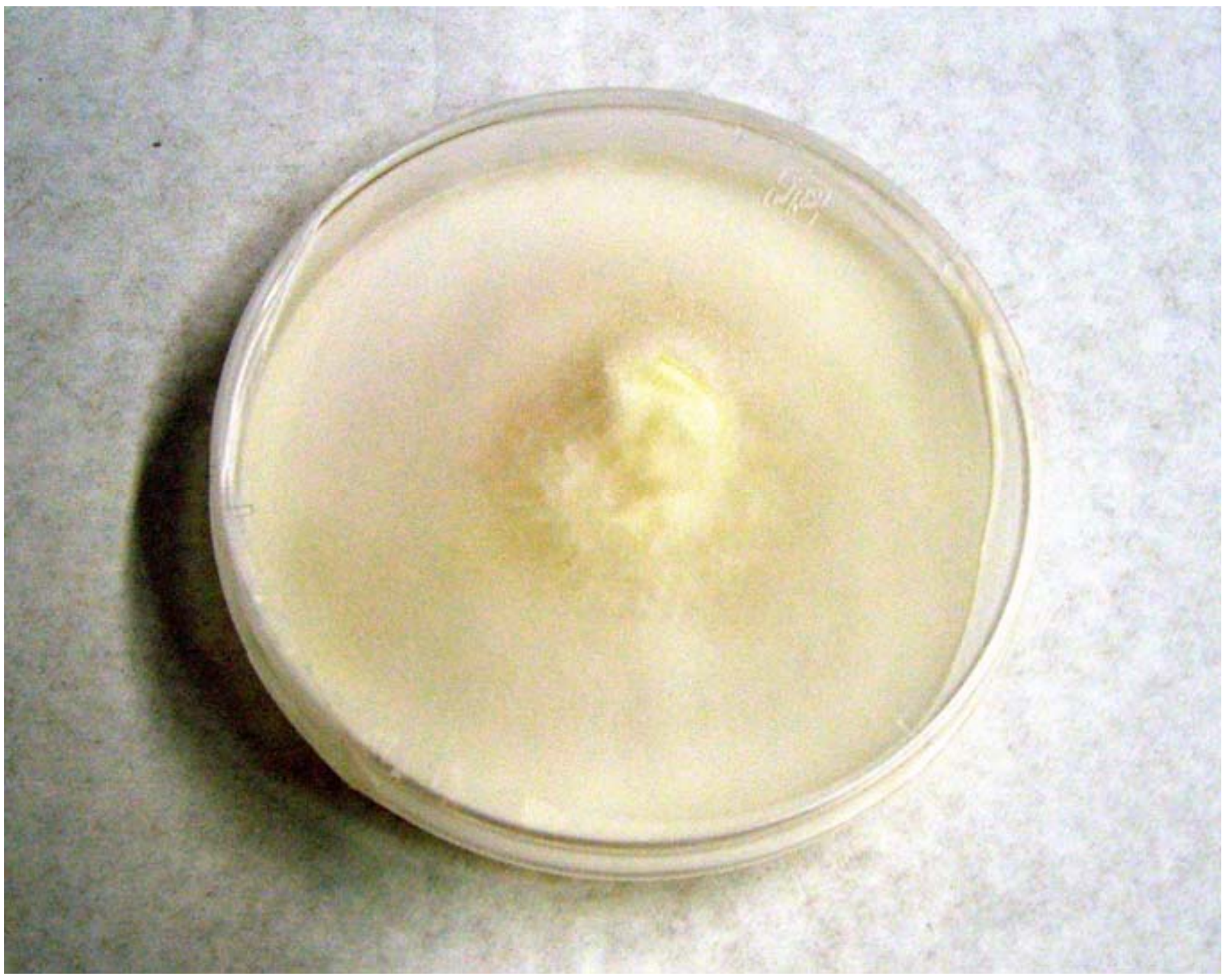

**Figure 3.** *T. pubescens* after 10 days on a Petri dish with MEA (4.8% w/v)

The genus *Trametes*, which belongs to the white-rot fungi, is assumed to be one of the main producers of laccases. Among them, *Trametes pubescens* has been described as a promising laccase producer [18] (Figure 3). In order to produce laccases, white-rot fungi have to be cultured under specific conditions. Two types of culture techniques are used: Solid-state fermentation (SSF) and Submerged fermentation (SmF) (Figure 4).

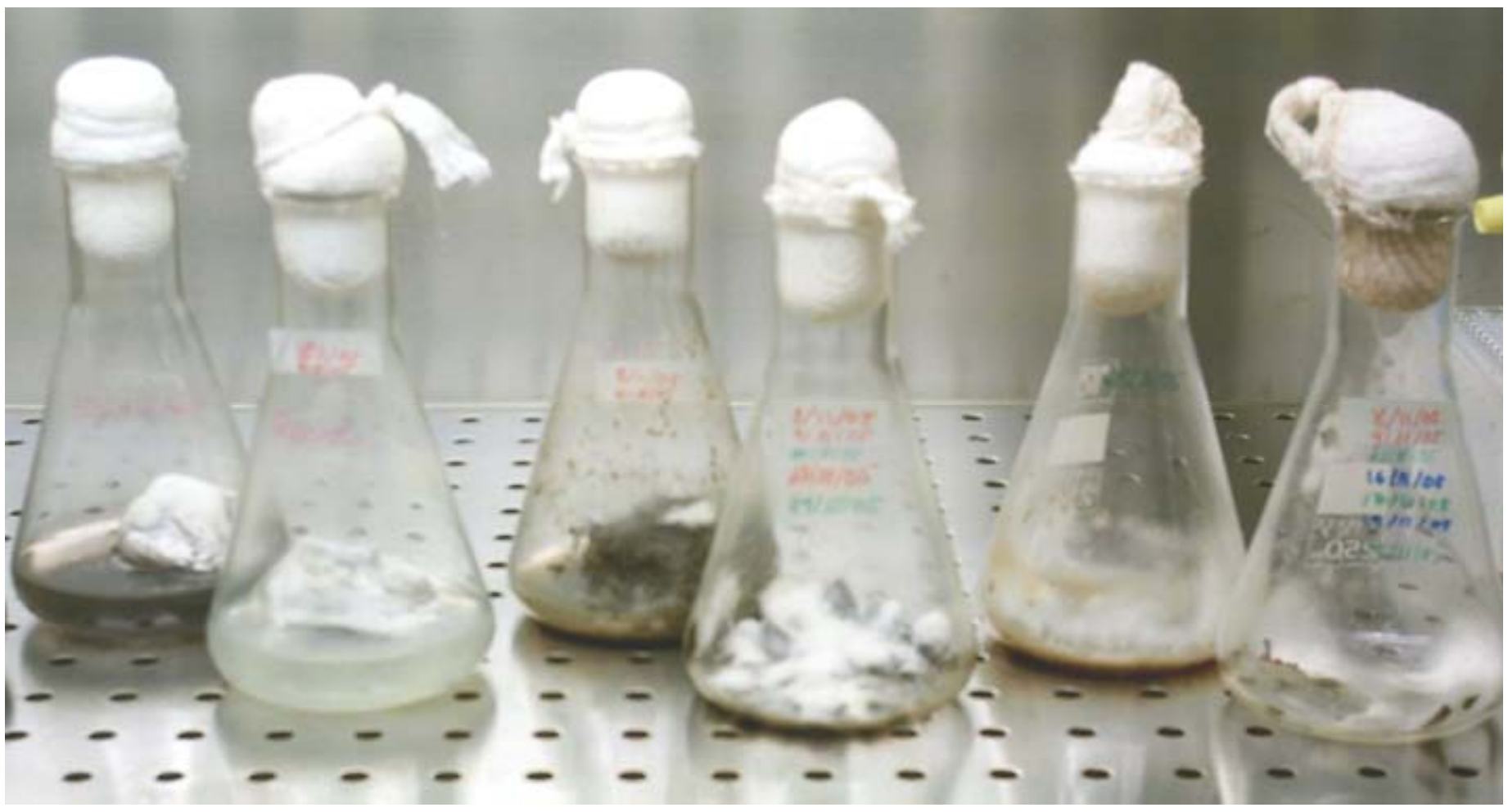

**Figure 4.** Cultures of *T. pubescens*: Left: SmF cultures; Right: SSF cultures

The SSF is defined as any fermentation process occurring in absence or near absence of free liquid, using an inert substrate or a natural substrate as a solid support [19]. The former works as an attachment place for the microorganism, whereas the latter also acts as a carbon source, which considerably reduces the production costs [20]. SSF is advantageous in obtaining concentrated metabolites and subsequent purification procedures are economical [21-22]. In SSF, the microorganisms grow under conditions close to their natural habitat. This may allow them to produce certain enzymes and metabolites, which usually would not be produced or would only be produced at a low yield in SmF [19]. Therefore, the selection of an adequate support is essential, since the success of the process depends on it. The most important factors to take into account are chemical composition, particle size, and of course cost and availability.

In recent years, there has been an increasing trend towards the utilization of organic wastes such as residues from the agricultural, forestry and alimentary industries as raw materials to produce value-added products by the SSF technique [23]. The use of such wastes, besides providing alternative substrates, helps to solve environmental problems, which are caused by their disposal. Furthermore, most of these wastes contain lignin or/and cellulose and hemicellulose, which act as inducers of the ligninolytic activities. Moreover, most of them are rich in sugars making the whole process more economical. All this makes agro-wastes very suitable as raw materials for the production of secondary metabolites of industrial significance by microorganisms. In particular, the present study focuses on laccase production by *T. pubescens*.

Banana skin has been selected because of its high content in carbohydrates, which due to their organic nature are easily metabolized by the fungus [24], and it has the necessary consistency to serve as a supporting material. In addition, its content in ascorbic acid exerts an inhibitory effect against bacteria [25]. Moreover, the production of banana is well spread around the world and main producers are grouped by regional organizations (Figure 5) lead by the Network for the Improvement of Banana and Plantain (INIBAP) [26]. The global banana production is estimated in 72.6 million tons per year [27] and the banana processing industry generates a huge amount of solid wastes, which are dumped in landfills, rivers, oceans and unregulated dumping grounds. Therefore, the reutilization of such wastes would help to diminish the pollution problems caused by their disposal.

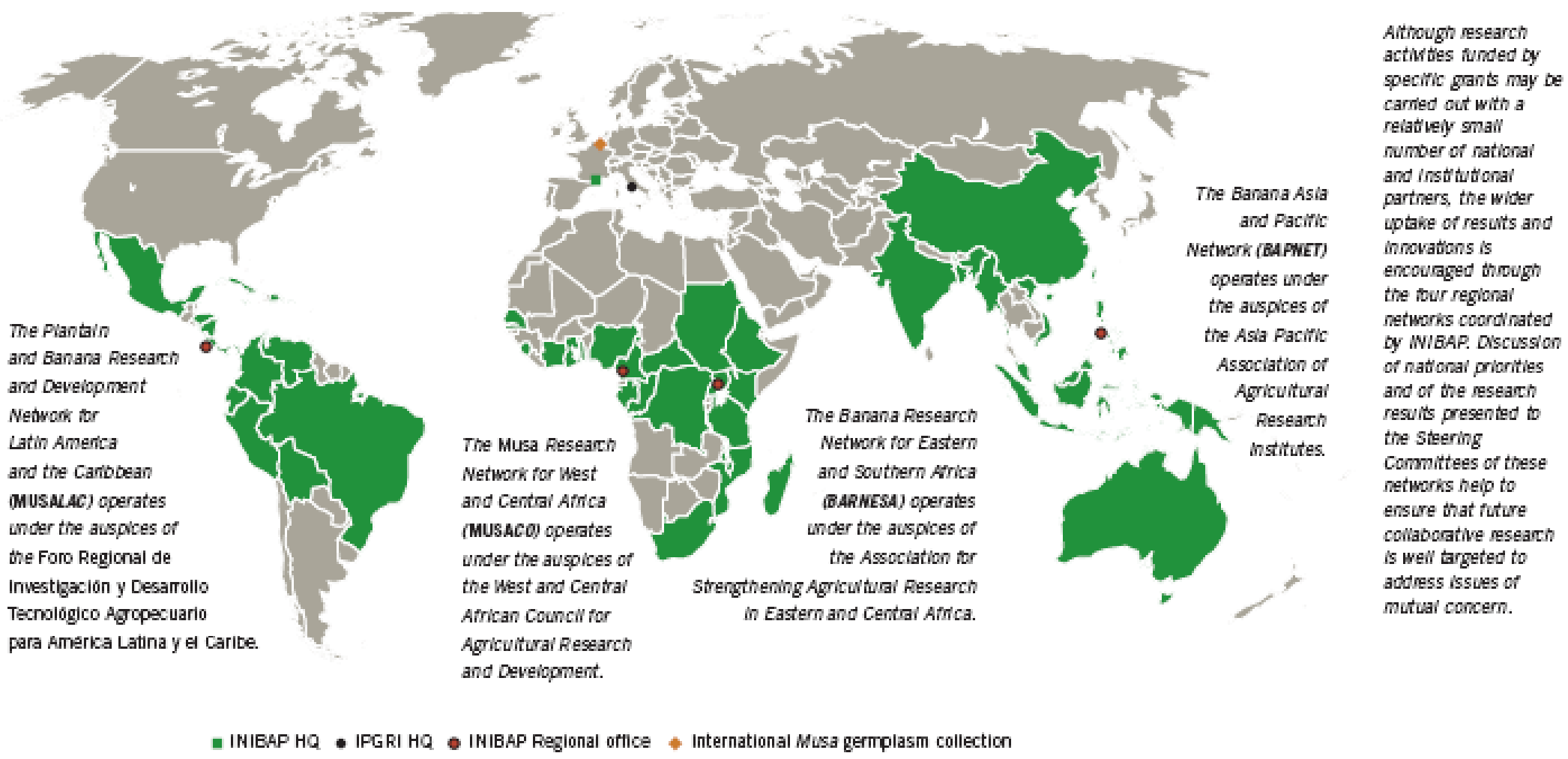

**Figure 5.** Main banana producers and regional organizations [26]

# 2 Formulation of the problem

The enzymatic treatment of wastewater requires the production of large amounts of enzymes, in this case laccases, at low cost. The current commercial price of laccases is high, constituting a drawback for its use. A low-cost process for the production of laccases is necessary for a sustainable enzymatic wastewater treatment. Therefore, it is necessary to establish an easy and low-cost procedure for the production of laccase.

# 3 Hypothesis

As previously commented in the introduction section, the banana skin is a good choice as an agro-waste material for the low-cost production of laccase by *T. pubescens* under SSF conditions.

# 4 Objectives

## 4.1 Main Objective

The main objective of the present wok is to investigate the potential of banana skin as a support-substrate for the production of laccase by *T. pubescens* under SSF conditions, since the use of such support would mean an important reduction in the production costs.

## 4.2 Sub objectives

- Compare the dye degradation capabilities of the produced laccase with a commercial laccase.
- Analyze the attachment of *T. pubescens* on the banana skin.

# 5 Methodology

## 5.1 Microorganism

*Trametes pubescens* (CBS 696.94) was maintained on 4.8 % (w/v) malt extract agar (MEA) plates at 4ºC and sub-cultured every three months (Figure 3).

## 5.2 Agro-waste material

Chopped banana (*Musa cavendishii*) skin (particle size 7.5 mm × 7.5 mm), purchased at a local store, was used as support-substrate for laccase production by *T. pubescens* under SSF conditions. Table 1 shows the composition of the banana skin.

Banana skins were pre-treated as follows: Each 10 g of fresh support was first soaked for an hour in 30 ml of 83.17 mM KOH to neutralize the organic acids [28]. Then, they were thoroughly washed with distilled water and dried at moderate temperature. Prior to use, the skins were autoclaved at 121ºC for 20 minutes.

**Table 1.** Chemical composition (% dry matter) of the banana skin [25]

| Compound (g per 100 g) | | Mineral and ascorbic acid content (mg per 100 g) | |
|---|---|---|---|
| Dry matter | 14.08 | Calcium | 7 |
| Crude protein | 7.87 | Sodium | 34 |
| Crude fat | 11.60 | Phosphorus | 40 |
| Crude fibre | 7.68 | Potassium | 44 |
| Total ash | 13.44 | Iron | 0.93 |
| Carbohydrates | 59.51 | Magnesium | 26 |
| Moisture | 78.4 | Sulphur | 12 |
| | | Ascorbic acid | 18 |

## 5.3 Culture conditions

The composition of the culture medium consisted of 2 g/L of glucose, 0.9 g/L of $(NH_4)_2SO_4$, 2 g/L of $KH_2PO_4$, 0.5 g/L of $MgSO_4$ $7H_2O$, 0.1 g/L of $CaCl_2$ $2H_2O$ and 0.5g/L of KCl in citrate-phosphate buffer (pH 4.5). This medium is identical to that used in Rodriguez Couto et al. 2006 [29], except for the amount of glucose, since banana skins also serve as a carbon source for the fungus.

Prior to use, the medium was autoclaved at 121ºC for 20 minutes and 0.5 g/L of thiamine sterilized by filtration (0.22 μm) was added subsequently. The cultures were performed in cotton-plugged Erlenmeyer flasks (250 ml) containing 7 g of chopped banana skins and 20 ml of culture medium. Inoculation was carried out directly in the Erlenmeyer flasks. Three agar plugs (diameter, 7 mm), from an actively growing fungus on MEA, per Erlenmeyer were used as inoculum. The Erlenmeyer flasks were incubated statically under an air atmosphere at 30ºC and in complete darkness.

## 5.4 Analytical determinations

*Laccase activity* was determined spectrophotometrically as described by Niku-Paavola et al. (1990) [30] using ABTS (2,2'-azino-di-[3-ethyl-benzo-thiazolin-sulphonate]) as a substrate. It is well known that fungal laccases, among other enzymes, oxidize ABTS (green-colored molecule) to the cation radical $ABTS\cdot^{+}$ (dark green-colored molecule) [31]. For the case of ABTS, the colorimetric changes can be determined by measuring the change in absorbance spectroscopy at the wavelength of 436 nm [31]. The change in absorbance ($\Delta E$) at a particular time interval ($\Delta t$) for a particular reaction can be calculated by the Lambert –Beer equation (1), where c is the concentration of the substrate in molar units, ε is the extinction coefficient in $M^{-1}$ $cm^{-1}$ and d is the path length of the sample the light beam traverses in cm. The extinction coefficient for the oxidation of ABTS at 436 nm is 29.3x$10^3$ $M^{-1}$ $cm^{-1}$ [32] and the path length of the optical cell used is 1 cm.

$$\frac{\Delta c}{\Delta t} = \frac{\Delta E / \Delta t}{\varepsilon \cdot d} = \frac{\Delta E / \Delta t}{29.3 \times 10^3 M^{-1} cm^{-1} \cdot 1cm} \quad [U/l] \tag{1}$$

The reaction was carried out directly in a 1.5ml cuvette at room temperature, containing 350µl of 20mM ABTS, and 1150µl of extracellular liquid diluted in 25mM succinic buffer (pH 4.5). The change in the absorbance was monitored for 2 minutes. Using Equation 1 and the dilutions of the cuvette and sample, the activity of the laccase can be calculated using Equation 2, and summarized in Equation 3.

$$Activity_{Lac} = \frac{\Delta E / 2\min}{29.3 \times 10^3 l\, mol^{-1}} \cdot \left( \frac{1500 \mu l}{1150 \mu l} \right) \cdot (Dilution\, of\ the\, sample) \quad [U/l] \tag{2}$$

$$Activity_{Lac} = 22.26 \cdot (Dilution\, of\ the\, sample) \cdot \Delta E \quad [U/l] \tag{3}$$

Where, one activity unit was defined as the amount of enzyme that oxidizes 1 µmol of ABTS per min.

## *5.5 Microscopic examination*

Banana skin samples were fixed with 6% glutaraldehyde in phosphate buffer 0.1 M for 4 h. Afterwards; the samples were washed twice with phosphate buffer 0.1 mM for 15 min at 4ºC. Then, the samples were postfixed with osmium tetroxide 0.1 M at 4ºC. Following fixation, the samples were washed twice with phosphate buffer 0.1 mM for 15 min. After that, the samples were dehydrated through an ethanol-amylacetate series to pure amylacetate and critical point dried using $CO_2$ as the transition liquid. The dried samples were mounted on aluminum stubs, sputter coated with gold (20 nm) and examined with a Jeol 6400 scanning electron microscope (SEM) at 15 kV, belonging to SRCiT (Scientific and Technical Services) of the Rovira i Virgili University (Tarragona, Spain).

### 5.6 Mathematical analysis of the SEM images

SEM images were analyzed using MATLAB 6.5 (MathWorks Inc., Natick, MA). The relief of the SEM images was obtained by a cross-section analysis of the gray-scale information. The color information of the images was treated in only one component*, since SEM images presented gray-scale information. This information was normalized between cero and one, being cero the darkest pixel of the image and one the brightest (Equation 4).

$$x(a,b) = \frac{p(a,b) - \min(P)}{\max(P) - \min(P)} \quad , \quad \forall \quad 1 \le a \le m \; and \; 1 \le b \le n \qquad (4)$$

where,

*n* is the number of pixels of one line of the SEM image

*m* is the number of pixels of one column of the SEM image

*p(a,b)* is the value of the pixel in the row a and column b

*P* is the matrix that contains the values of all the pixels of the SEM image

*x(a,b)* is the normalized gray-scale value of the pixel in the row a and column b

Sections of 50µm x 1µm (measured using the scale of the image in pixels) were selected and compressed to a relation of 1µm per pixel. Afterwards, Discrete Fourier Transformation (DFT) was applied to the cross lines of SEM images in order to obtain the frequency information according to Equation 5.

$$X(k) = \sum_{n=1}^{N} x(n) e^{(-2j\pi(k)(n-1)/N)} \quad , \quad \forall \quad 0 \le k < \frac{N}{2} \qquad (5)$$

* RGB: (Red,Blue,Green) is a common 3 component format for storage the color information of an image. Gray-scale images contain the same information in each component of color.

where,

$N$ is the number of pixels of one section of the SEM images

$j$ is the imaginary unit

$x(n)$ is the normalized gray-scale value of the $n^{th}$-pixel of the section

$X(k)$ is the $k^{th}$ harmonic component of the signal

The magnitude of each component $|X(k)|$ can be understood as the amount of structures of a certain size in the image. The DFT is based on the interpolation of harmonic sinusoidal signals to represent a continuous signal. The different harmonics, (Figure 6), have intrinsic information about structures of the same size* in the images. If the third component of the DFT has a large magnitude, it can be assumed that there are many structures of $N/3$ size. Figure 7a shows an example of a cross-section image and its harmonic composition while Figure 7b represents the magnitude of each harmonic component $X(k)$.

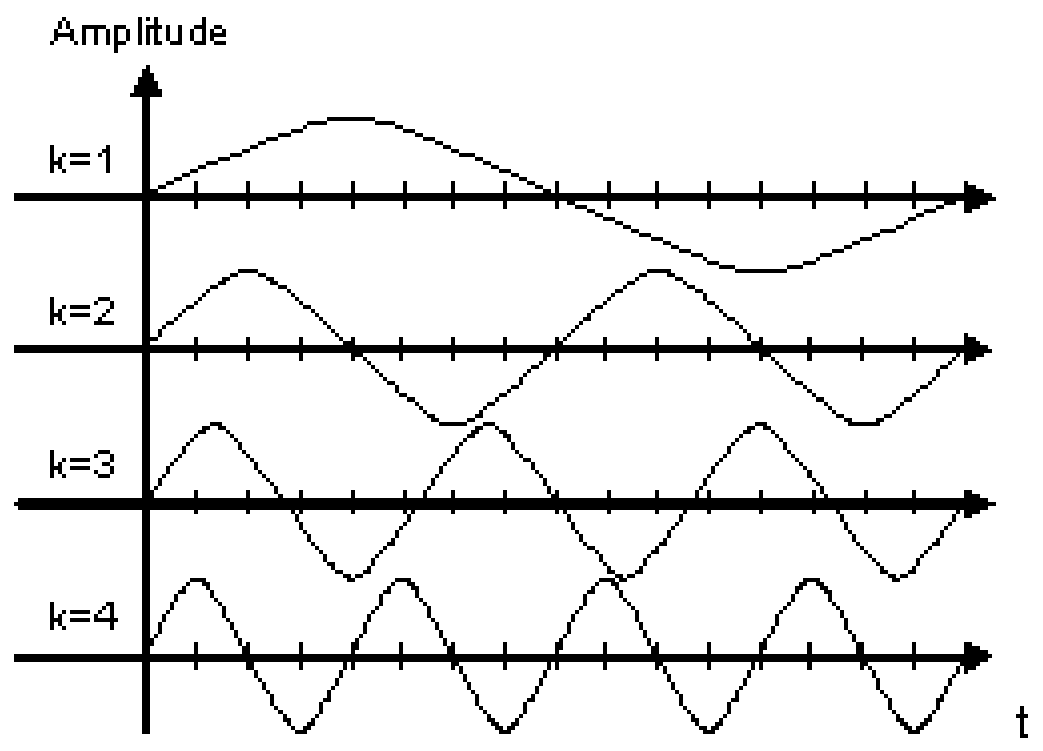

**Figure 6.** Sinusoidal harmonic signals with same amplitude

* Size: Analogy of time period of a sinusoidal signal with the number of pixels of a certain structure in an image.

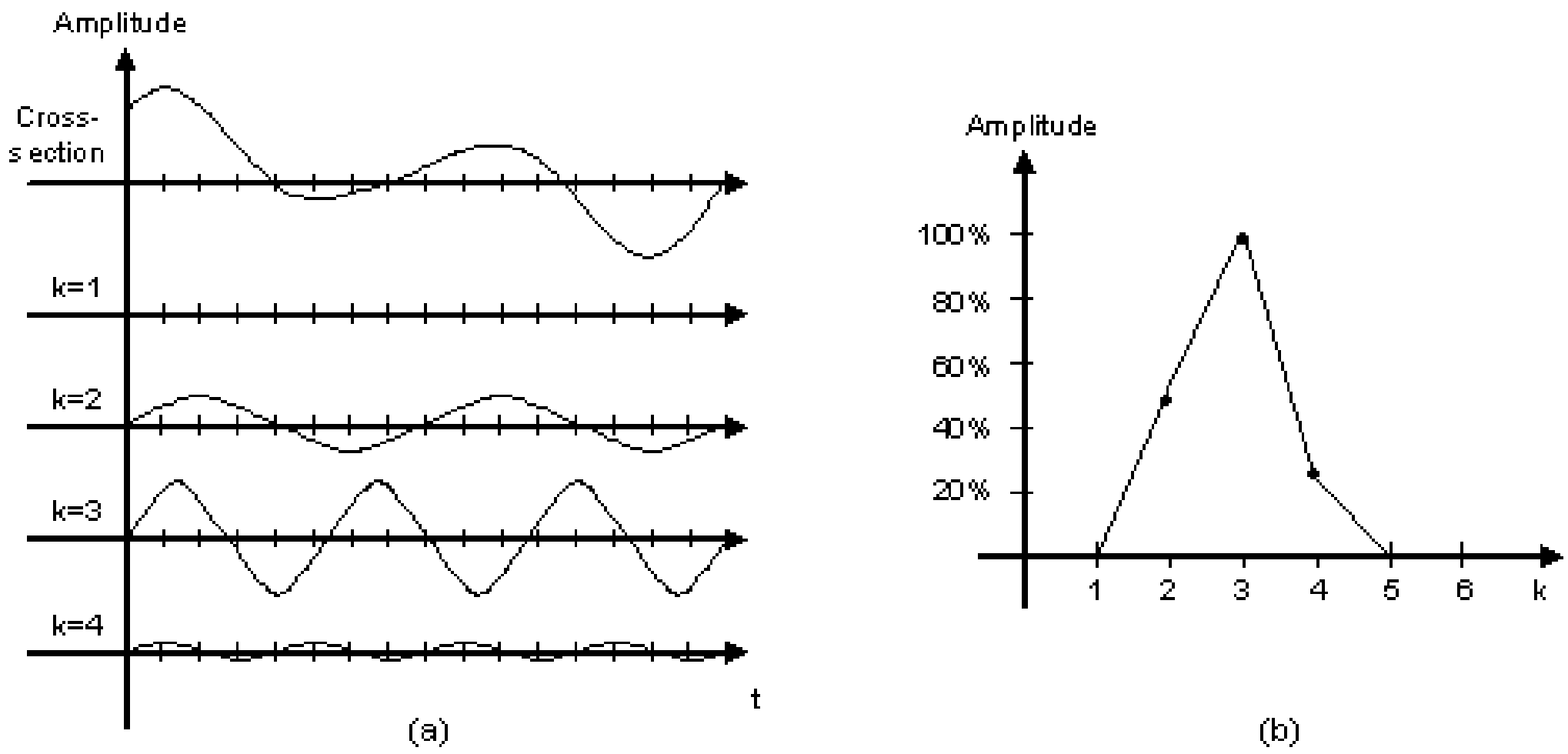

**Figure 7.** Example of the DFT of a cross-section signal. (a) Cross-section signal and harmonic components, (b) Magnitudes of the harmonic components *X(k)*

## 5.7 *Decoloration studies*

In order to determine the dye degradation capability of the produced laccase two dyes were used, Remazol Brilliant Blue R (RBBR), purchased from Sigma Aldrich (St. Louis, MO, USA) and Methyl Green (MG), purchased from Merck (Germany). The characteristics of the dyes are summarized in Table 2. Stock solutions (0.1% w/v in water) were stored in the dark at room temperature.

Culture broth from banana-skin cultures of *T. pubescens*, collected on day 14, and a commercial laccase from *T. hirsuta*, supplied from Novo Nordisk (Denmark), were used for dye decoloration experiments.

The reaction was carried out directly in the spectrophotometer cuvette and the reaction mixture (final volume 1.5 ml) consisted of an aqueous solution of dye and extracellular liquid or commercial laccase (300 U/l, final concentration) in succinic buffer (pH 4.5). Dye concentrations were selected in order to obtain one unit of absorbance at the maximum wavelength in the visible spectrum (0.133 g/l for RBBR and 0.033 g/l for MG, final concentration). All the reactions were incubated at room temperature, in static conditions and in complete darkness.

The residual dye concentration was measured spectrophotometrically and associated with the decrease in the absorbance at the peak of maximum visible wavelength (595 nm for RBBR and 630 nm for MG). Dye decoloration was expressed in terms of percentage. A control test containing the same amount of a heat-denatured laccase was also performed in parallel. The assays were done twice, the experimental error being below 10%.

# 6 Results and discussion

## *6.1 Laccase production*

As shown in Figure 8, laccase production began on the 3rd day (63 U/l) and, then, it sharply increased up to a maximum activity of nearly 1600 U/l at the end of cultivation. A good reproducibility of the enzyme production can be noticed. Also, the smooth increase of the enzyme activity (absence of short-term peaks) easies the collection of the medium, that contains the laccase, since a difference of one day is not critical.

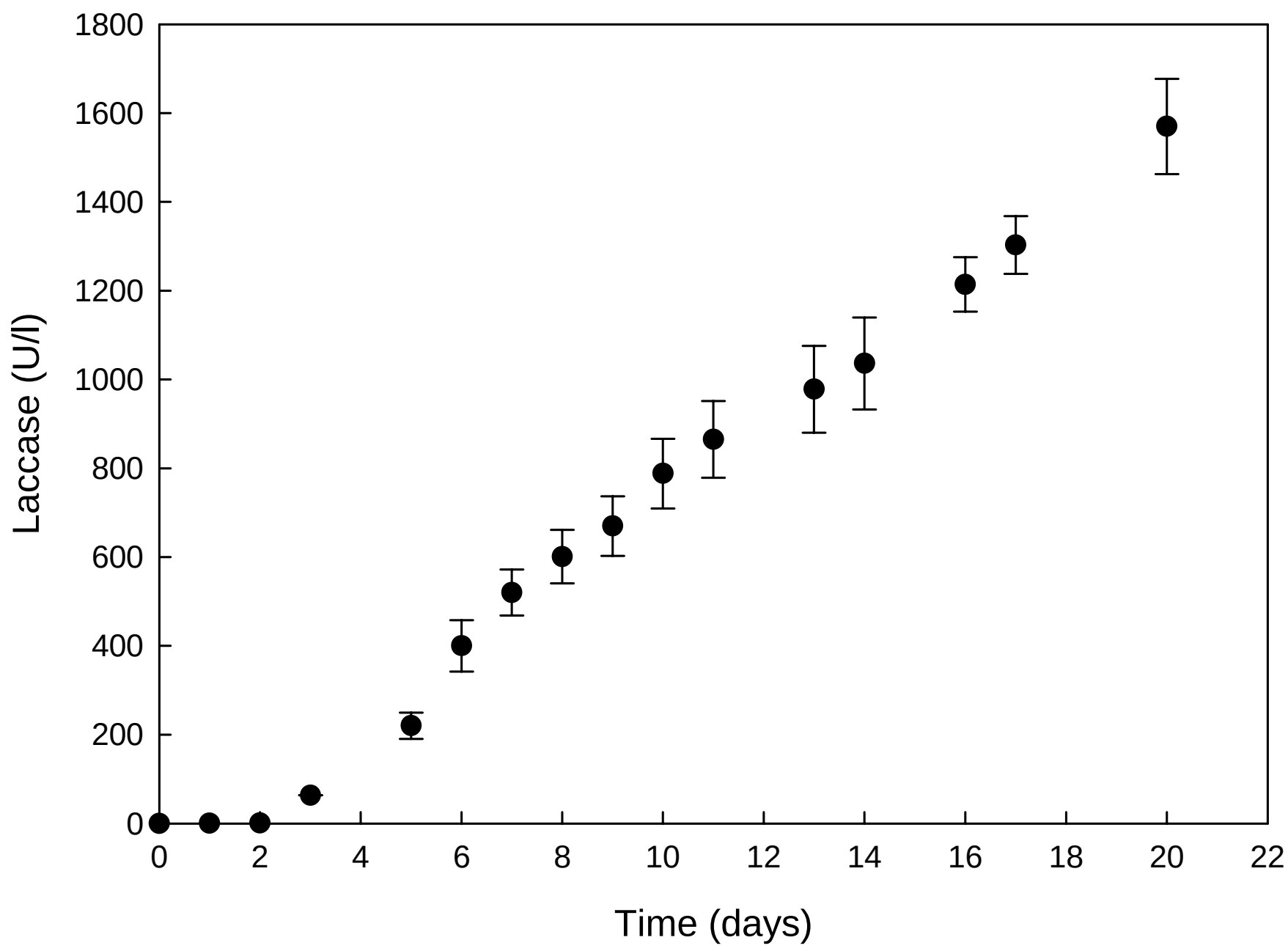

**Figure 8.** Laccase production by solid-state cultures of *T. pubescens* grown on banana skins

Another important factor was the attachment of the fungus to the banana skin, making easier the collection of the medium. This is very interesting for the subsequent application of this enzymatic complex to biotechnological processes, since the purification stage would be more economical.

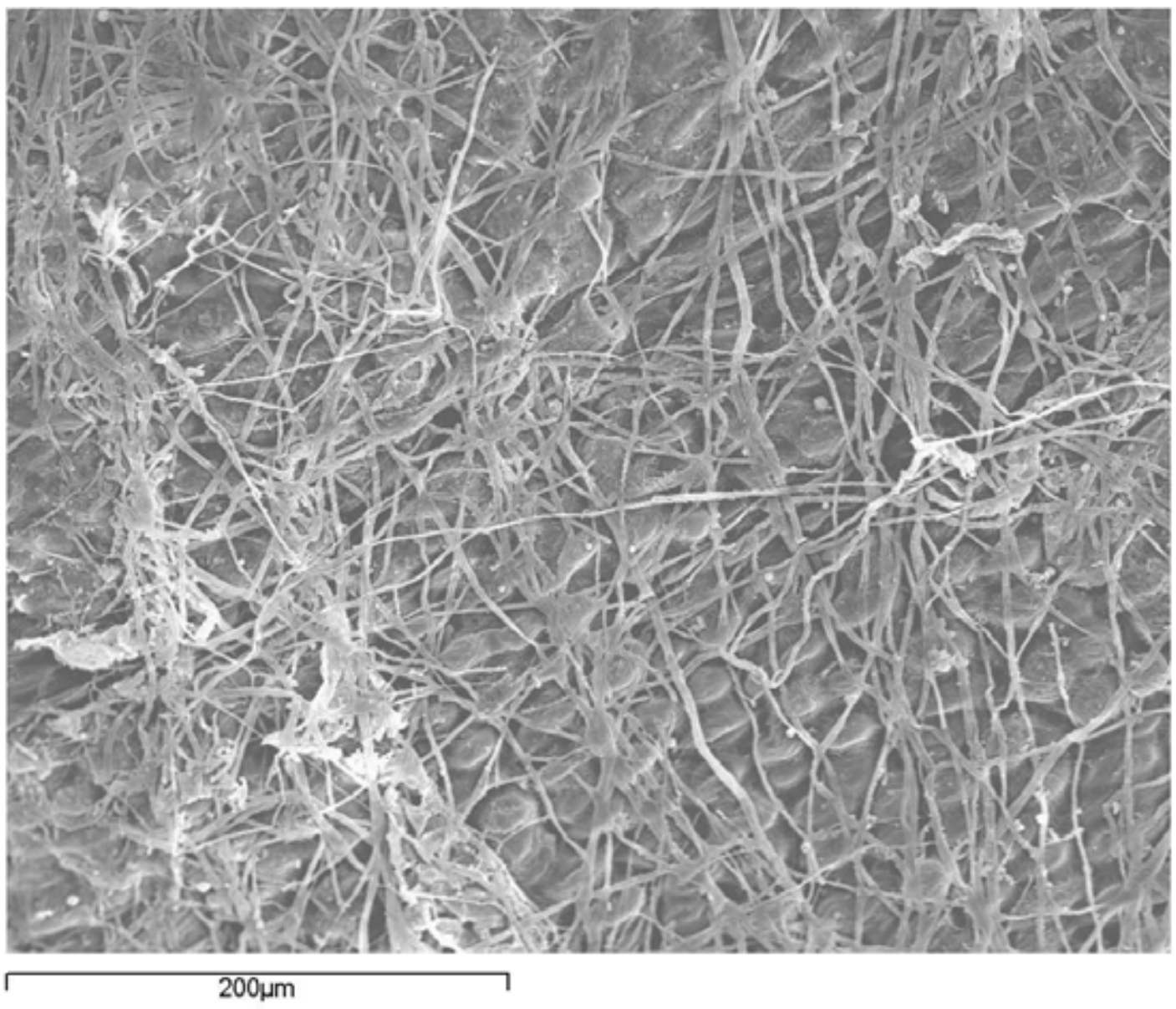

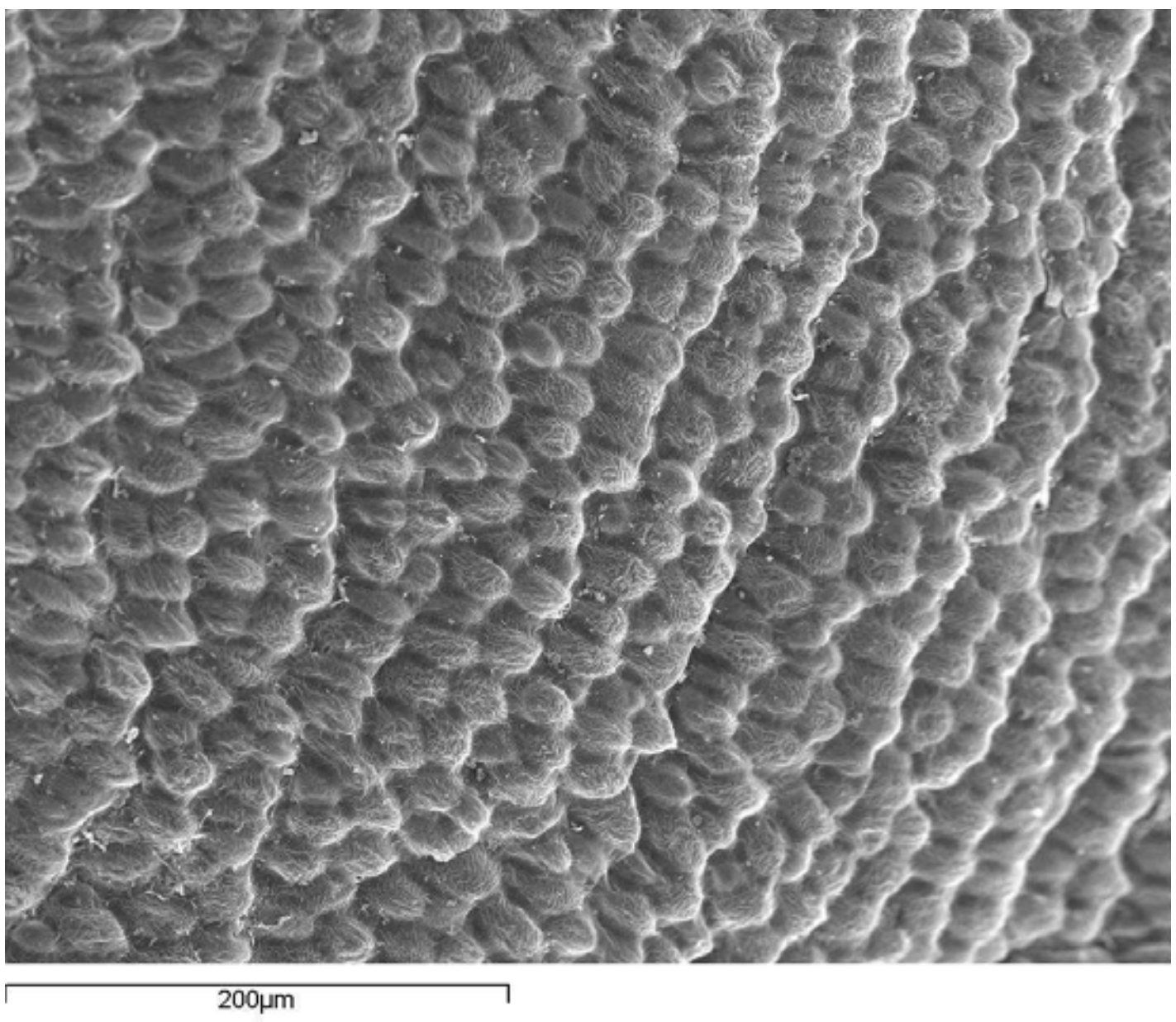

**Figure 9.** SEM microphotographs of banana skin: Up: with fungus; Down: without fungus

## 6.2 Microscopic examination

The most important characteristics that influence adhesive behavior of filamentous fungi to the support are hydrophobicity and surface charge. Figure 9 shows SEM microphotographs of banana skin with and without fungus. It can be observed that the fungus grew well attached to the banana skin. This is due to the high hydrophobicity of the banana skin, which eases the attachment of the fungus to the carrier [33]. Therefore, banana skin is very suitable as an attachment place for filamentous fungi. This together with its high content in carbohydrates (Table 1) makes banana skin very suitable as a support-substrate to perform solid-state processes.

SEM images were analyzed by software techniques in order to obtain information about the position of the fungus on the banana skin cells. Dotted line in Figure 10 represents a cross-section segment of the banana skin showing an expected half-sinusoidal shape for each cell. Higher values correspond to the upper part of the cells; meanwhile, lower values correspond to the space between cells. Solid line represents the cross section of the banana skin with fungus on it. Higher frequency peaks can be noticed as an indicator of the tubular structure of the fungus. Also, big covered areas show places were the fungus agglomerates. The growth of the fungus is not regular and does not cover homogenously the surface of the cells; it is concentrated in different places where carbon source is surely more accessible. It also presents tubular structures (hyphae) that form clusters and grow irregularly, not following a specific pattern.

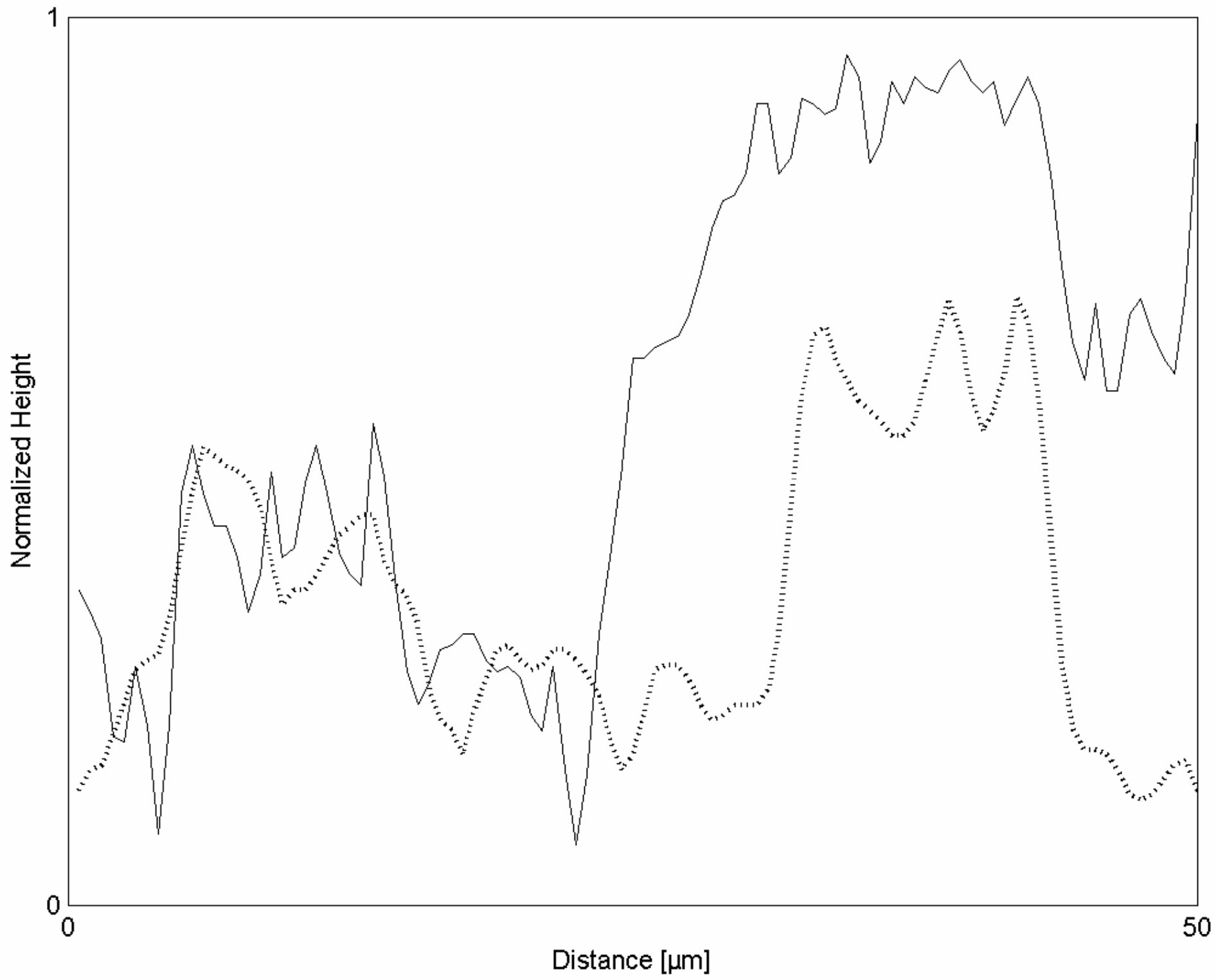

**Figure 10.** Cross sections of banana skin with (solid line) and without fungus (dotted line)

Figure 11 presents a Fourier frequency analysis of the same cross-section images, where lower frequency peaks correspond to the size of the cells, and high frequency peaks correspond to their roughness. Fourier analysis shows the mean size of the banana skin cells through the vertical and the horizontal axes (17 and 25µm, respectively).

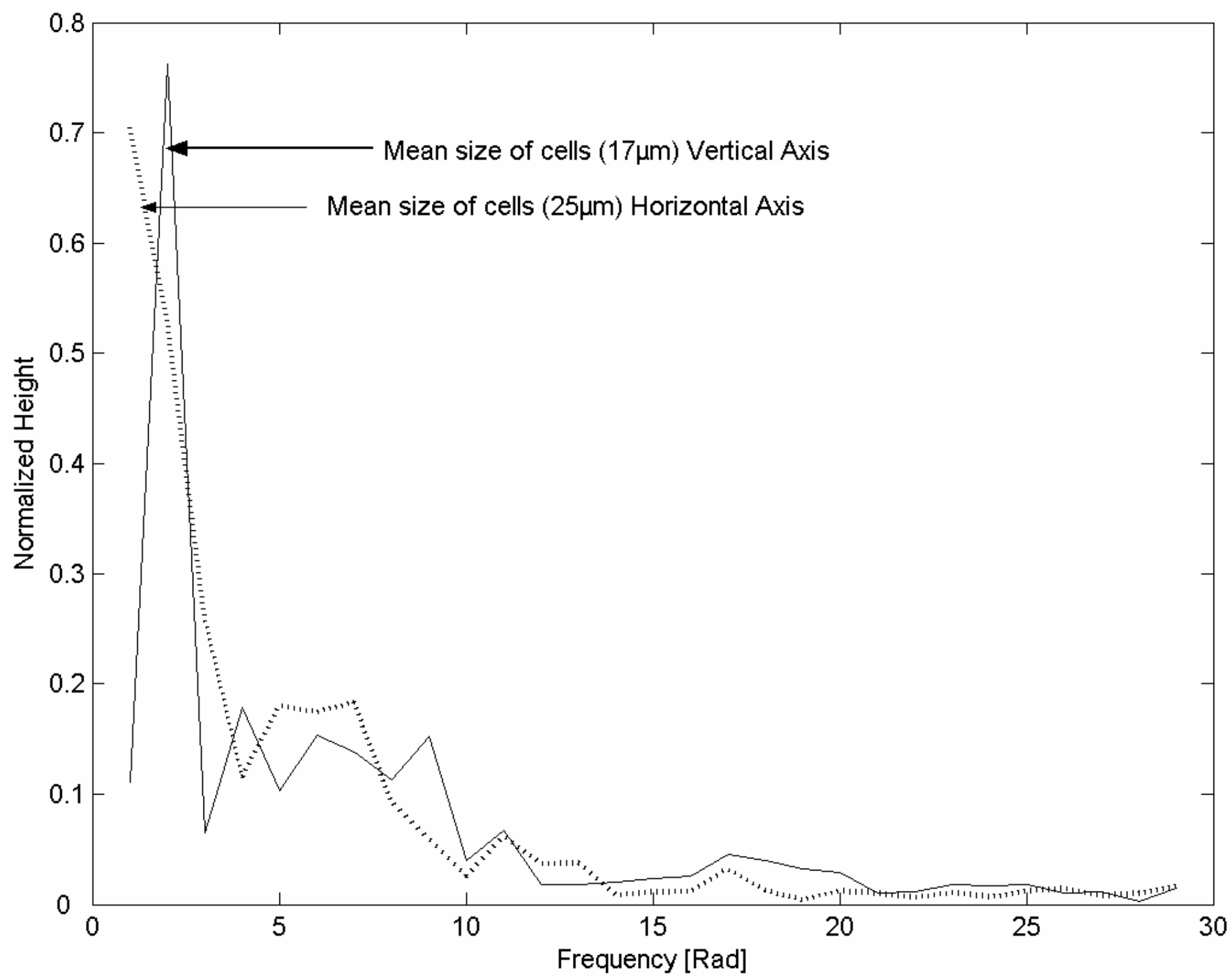

**Figure 11.** Fourier frequency analysis of the cross-section of the banana skin with fungus. Vertical axis analysis (solid line) and horizontal axis analysis (dotted line). The two peaks, 25 and 17 µm, correspond to the second and third harmonics, respectively, of the Discrete Fourier Transformation (DFT) with N = 50 (1pixel = 1µm) extracted from information in Figure 10 (cross-section of banana skin with fungus)

## *6.3 Decoloration studies*

The ability of white-rot fungi to decolorize synthetic dyes has been widely studied, particularly with *Phanerochaete chrysosporium* and *Trametes versicolor* [34]. In the present study, we assessed the ability of the extracellular fluid from *T. pubescens*, a little studied white-rot fungus, to decolorize two structurally different synthetic dyes (RBBR and MG). The decoloration of type model dyes is a simple method to assess the aromatic degrading capability of ligninolytic enzymes [35].

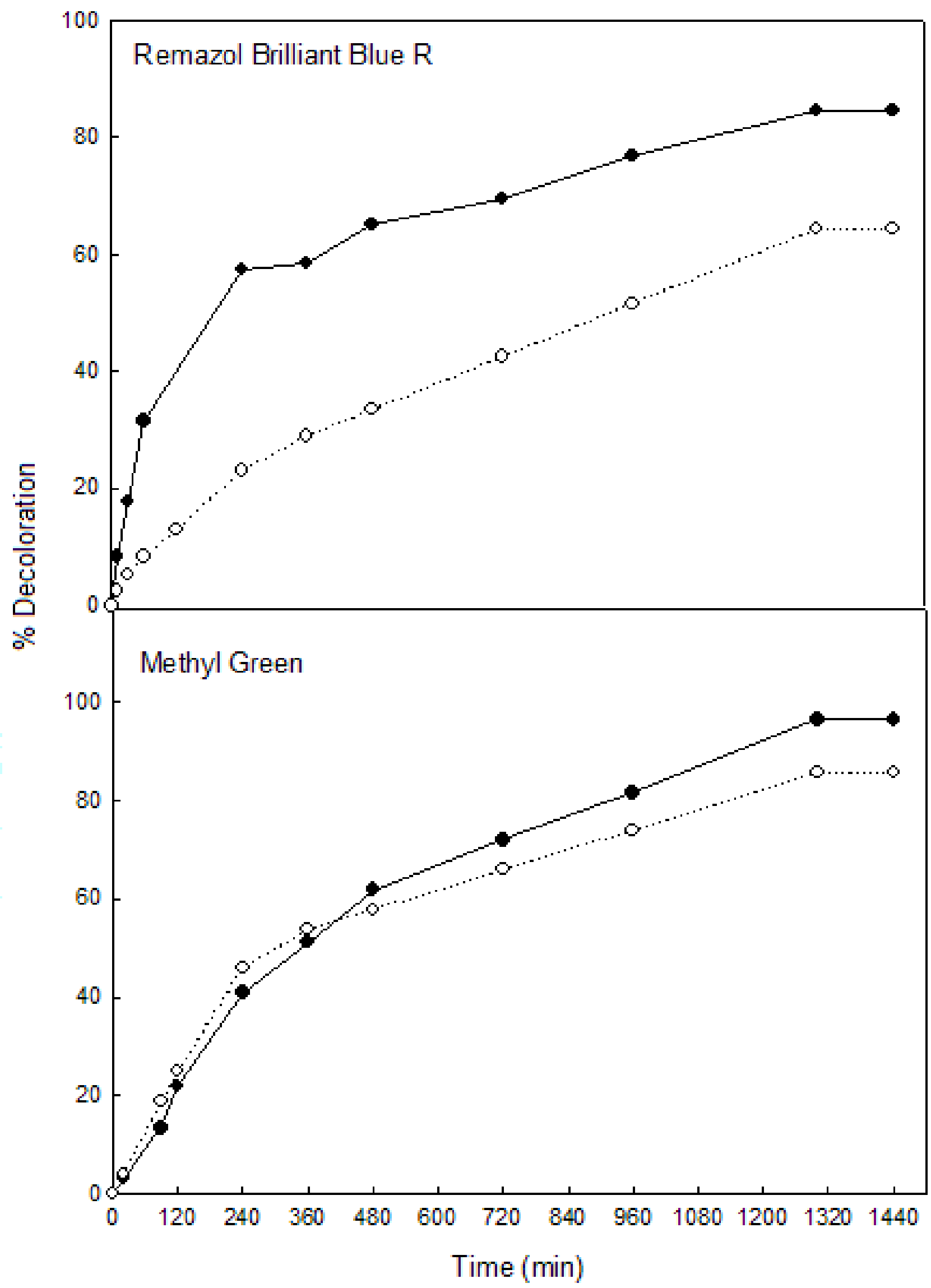

**Figure 12.** Profile of dye decoloration attained: (●) extracellular liquid from banana skin cultures of *T. pubescens*; (o) commercial laccase

As it can be seen in Figure 12, the decoloration rate obtained was very different in each case. After 4 h of treatment, RBBR showed a degree of decoloration of about 57 %. However, from here onwards, decoloration proceeded very slowly and it reached a value of 84.5 % in 21 h. This could be due to either enzyme inhibition by some products generated in the decoloration process or substrate inhibition. These results differ from those found by Soares et al. (2001) [36], who reported that the addition of a redox mediator was necessary for RBBR decoloration by a commercial laccase from Novo Nordisk, Denmark (produced by SmF of a genetically modified *Aspergillus* microorganism). The discrepancy between our results and those from Soares et al (2001) [36] could be due to the difference in fungal species from which the laccase was obtained, the different culture medium and/or the technique used. In addition, the redox potential of laccases varies depending on the laccase source [37], which could also dictate the need of a redox mediator for the decoloration of a particular dye to occur.

MG was decolorized about 41 % in 4 h reaching a decoloration of 96.4 % after 21 h. Therefore, the decoloration rate was slower than that of RBBR but the final decoloration was higher (Figure 12). It has been reported that highly substituted triphenylmethane dyes required longer time to be decolorized or could only be decolorized to a certain extent [38].

When a commercial laccase was used, RBBR decoloration was significantly lower than that attained by laccase from banana skin cultures of *T. pubescens*. On the other hand, MG decoloration was very similar for both laccases (Figure 12).

It was observed that from 24 h of incubation onwards decoloration did not increase (data not shown). This could be due to enzyme inhibition by some products generated in the decoloration process.

Since equal doses (300 U/l) of laccases were used in the decoloration process, the difference in the decoloration efficiency of the two laccases was most likely due to the difference in laccase isoenzymes produced by the different strains as well as to the difference in specificities to different dyes of diverse structures [39]. In addition, as commented above, it could also be due to the difference in the redox potential of laccases from different microorganisms.

The fact that *T. pubescens* was able to decolorize the dye RBBR with no mediator addition is very interesting, since this dye is frequently used as a starting material in the production of polymeric dyes. Also, it represents an important class of often toxic and recalcitrant organopollutants [40].

## 7 Conclusions

On one hand, the results clearly showed the enormous potential of banana skin as a support-substrate for production of laccase at low cost by *T. pubescens* under solid-state conditions. In addition, the laccase produced presented a highly decolorizing ability, especially for anthraquinonic dyes. This makes laccase from this fungus very attractive for further investigations as well as for its application to different biotechnology areas.

On the other hand, banana is a highly produced fruit around the world, and the skin, its main waste. These circumstances make the banana skin a proper material for a low-cost production of laccases by reducing the use of other carbohydrate sources as glucose, frequently used in this type of processes, and thus the cost of the production. Another important advantage of this support-substrate is the good attachment with the fungus making easier the separation of the liquid media, which contains the enzyme, from the culture. In general, the laccase produced by *T. pubescens* grown on banana skin under SSF conditions, presented high decoloration ability.

## 9 Publications